\documentstyle[epsf,12pt]{article}
\begin{document}
\begin{center}
\Large{\bf The search for strongly decaying exotic matter}\\
\large{R.S. Longacre$^a$\\
$^a$Brookhaven National Laboratory, Upton, NY 11973, USA}
\end{center}
 
\begin{abstract}
In this paper we explore the possibility of detecting strongly decaying 
exotic states. The dibaryon(2.15) $J^P$ = $2^+$ state which decays into 
d $\pi$ is the example we use in this report.
\end{abstract}
 
\section{Introduction} 

At the STAR experiment we can collect hundreds of million ultra-relativistic 
heavy ion collisions. Light nuclei and anti-nuclei emerge from these collisions
during the last stage of the collision process. The quantum wave functions of 
the constituent nucleons or anti-nucleons, if close enough in momentum and 
coordinate space, will overlap to produce composite systems. The production 
rate for the systems with baryon or anti-baryon B is proportional to the baryon
or anti-baryon density in momentum and coordinate space, raised to the power 
$\vert$B$\vert$, and therefore exhibits exponential behavior as a function of
$\vert$B$\vert$. Figure 1 shows the exponential\cite{Armstrong} invariant 
yields versus baryon number in $\sqrt{s_{NN}}$=200 GeV central Au+Au 
collisions. Empirically, the production rate reduces by a factor of 1.1 x 
$10^3$(1.6 x $10^3$) for each additional nucleon (anti-nucleon) added. The 
measurement of hundreds of million of events make it possible to probe up to a 
scale of five in baryon number. The baryon four data points come from a STAR
measurement published in Ref.\cite{He4}. It should be noted that there are no 
baryon five nuclear fragments that live long enough or decay weakly such that
they would have a displaced vertex\cite{He4}.

The paper is organized in the following manner:

Sec. 1 explores nuclear states that have been measured.

Sec. 2 explores the possibility of detecting strongly decaying exotic states. 
The dibaryon(2.15) $J^P$ = $2^+$ state which decays into d $\pi$ is considered.

\begin{figure}
\begin{center}
\mbox{
   \epsfysize 4.8in
   \epsfbox{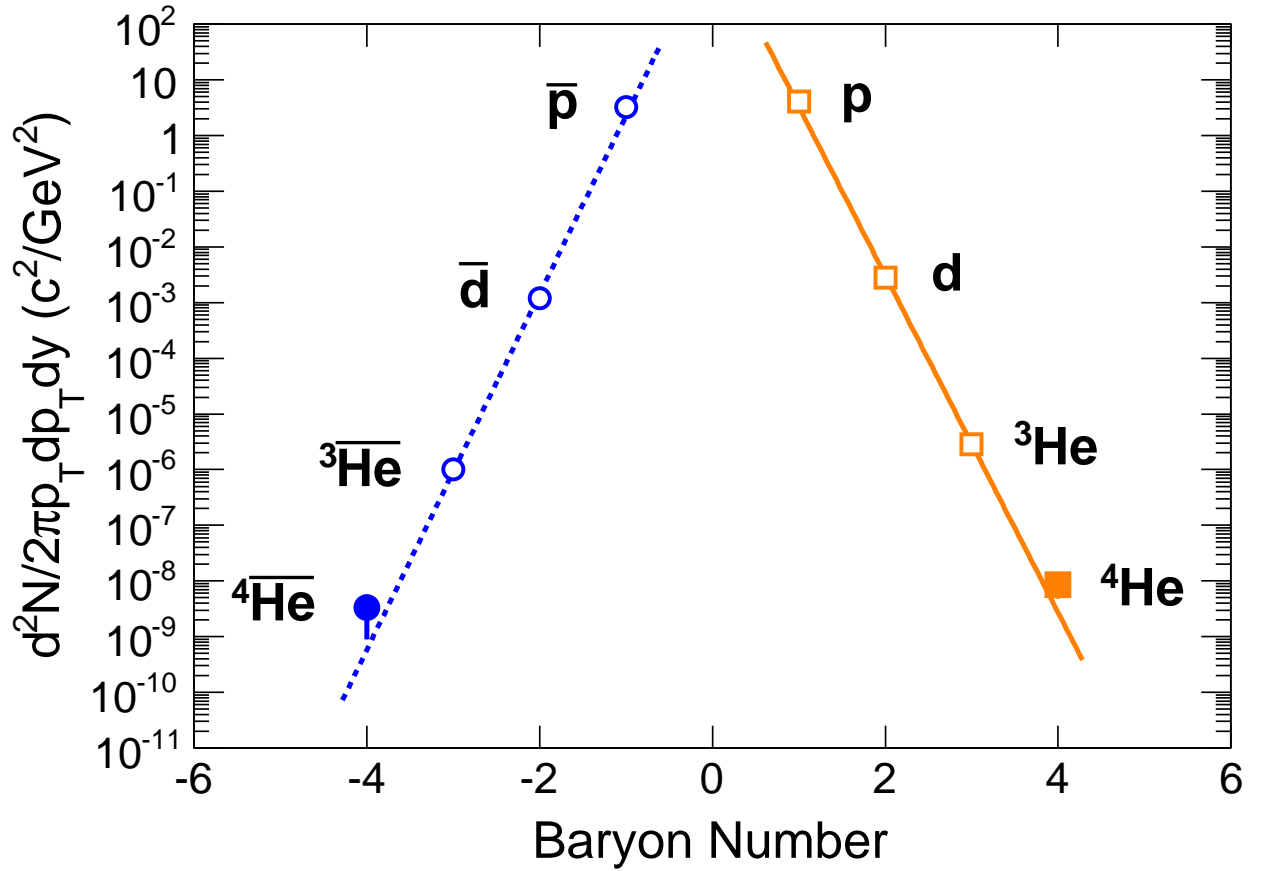}}
\end{center}
\vspace{2pt}
\caption{Differential invariant yields as a function of baryon number B, 
evaluated at $p_T$/$\vert$B$\vert$ = 0.875 GeV/c, in central 
$\sqrt{s_{NN}}$=200 GeV Au+Au collisions. Yields for tritons $ ^3H$ 
(anti-tritons $\overline{ ^3H}$) lie close to the position for $ ^3He$ and 
$\overline{ ^3He}$. The lines represent fits with the exponential formula 
$\alpha$ $e^{-r\vert B \vert}$ for positive and negative particles separately, 
where $r$ is the production reduction factor.}
\label{fig1}
\end{figure}

\section{Exotic States through strong decay.}

In the above section the states decayed by the weak interaction. The 
possibility of detecting strongly decaying exotic states is considered using 
the dibaryon(2.15) $J^P$ = $2^+$ state which decays into d $\pi$ as an example.
In the QGP(Quark Gluon Plasma) six quarks or anti-quarks could come together to
form a deuteron or anti-deuteron. However such states are loosely bound
and easily destroyed in the hadronic phase. The cross section for  $d$ $\pi$
scattering is 240 mb. This implies that the deuteron can only be formed in 
the final freeze-out of the hadronic system. At the time of freeze-out
many hadrons scatter and coalesce into compound or excited 
states(see Figure 2).
 
The dibaryon state interacts in three two-body scattering channels. Its mass 
is 2.15 GeV and has a strong interaction resonance decay width of 100 MeV. It 
interacts in the $N$ $N$ d-wave spin anti-aligned\cite{Arndt}, $d$ $\pi$ p-wave
spin aligned\cite{Oh}, and $\Delta$ $N$ s-wave spin aligned\cite{Schiff}. The 
dibaryon system mainly resonates in the s-wave $\Delta$ $N$ mode with a pion
rotating in a p-wave about a spin aligned $N$ $N$ system which forms a isospin
singlet. The pion moves back and forth forming $\Delta$ states with one nucleon
and then the other(see Figure 3). All three isospin states of the pion can be 
achieved in this resonance. Thus we can have $\pi^+$ $d$, $\pi^0$ $d$, and 
$\pi^-$ $d$ states. If the pion is absorbed by any of the nucleons it under 
goes a spin flip producing a d-wave $N$ $N$ system. The resonance decays into 
$N$ $N$, $\pi$ $d$, or $\pi$ $N$ $N$. In a meson system an analogous 
resonance is formed where a pion is orbiting in a p-wave about a 
$K \overline{K}$ in a s-wave\cite{Longacre}(see Figure 4). Both systems have 
a similar lifetime or width of $\sim$ .100 GeV.  

\begin{figure}
\begin{center}
\mbox{
   \epsfysize 6.0in
   \epsfbox{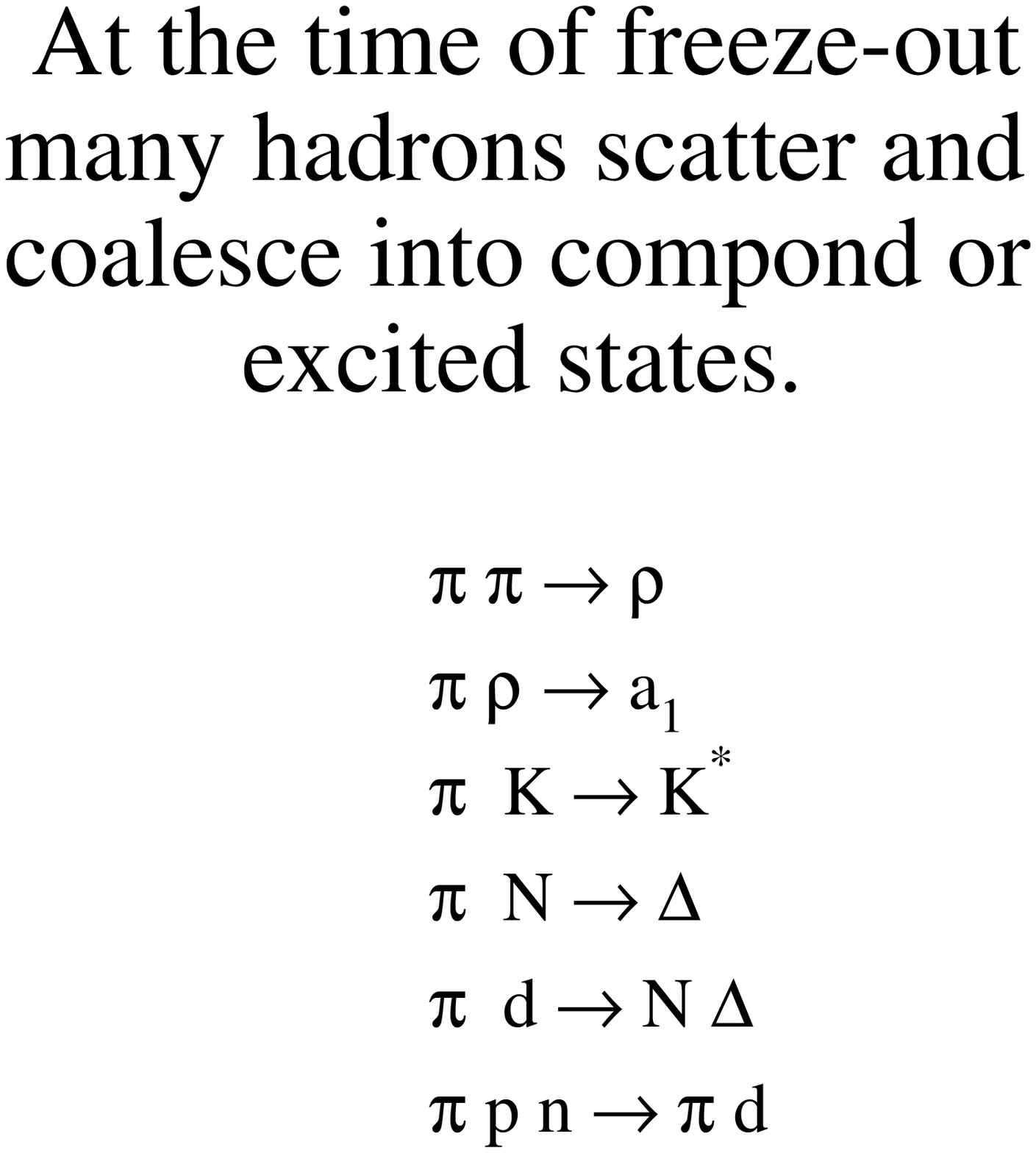}}
\end{center}
\vspace{2pt}
\caption{Above are a few of the compound or excited states that will form 
during the last stage of hadronic freeze-out.}
\label{fig2}
\end{figure}

In order to predict the rate for dibaryon production we turn to a Monte Carlo
heavy ion event generator\cite{Klaus}. This generator was a cradle to grave
going from initial partons to final state hadrons. Figure 5 shows the time line
in the center-of-mass frame for partons, then pre-hadrons and final hadrons. 
What is happening in the early times of the collision is of no importance 
for dibaryon formation, while the conditions of the hadrons at later times 
will determine the dibaryon production. For $\sqrt{s_{NN}}$=200 GeV central 
Au+Au collisions the spectrum is well measured. Therefore we can start the 
Monte Carlo at the intermediate times with a fireball of excited hadrons and 
let it evolve to the final state.

We start with an expanding cylinder of radius 10.0 Fermi filled with excited
hadrons with density and $p_t$ distribution that reproduces the 
$\sqrt{s_{NN}}$=200 GeV central Au+Au collisions. Figure 6 gives the measured
Au+Au spectrum which we will tune for. Mesons in the fireball cascade include
$\pi$, $K$, $\rho$, $\omega$, $a_1$, $\eta$, $\eta\prime$, $\phi$, $K^*$, 
$K^*(1420)$, $f_0(975)$, $a_0(980)$, $f_2(1270)$, $a_2(1320)$, $h_1(1170)$, 
$\rho(1700)$, $f_2(1800)$, $b_1(1235)$ and $f_2(1525)$. The cross section for 
$\pi \pi$ and $\pi K$ was determined from S-matrix phase shifts, while for 
$K \overline K$ we used the production of $\phi$, $f_0(975)$, $a_0(980)$, 
$f_2(1270)$, $a_2(1320)$,and $f_2(1525)$(see Figure 7). For $\rho \pi$
cross sections we used the production of $h_1(1170)$, $a_1(1260)$, and
$a_2(1320)$, while for $K^* \pi$ we used the production of $K^*(1420)$.
For $a_1 \pi$ we used $\rho(1700)$ and $f_2(1800)$(see Figure 8). Finally
for $\omega \pi$ we used $b_1(1235)$ and $\eta \pi$ we used $a_0(980)$ and 
$a_2(1320)$. For all other meson meson cross section we used the additive 
quark model. The particles produced from such scatterings were determined by 
a multi-pomeron chain model using a Field-Feymann algorithm(see Figure 9). 

Since we are detecting baryons and anti-baryons the $N N$, $N \pi$, $N K$, and
$\Delta N$ cross section and scattering ratios are obtained from data and
extracted S-matrix amplitudes(see Figure 10). All other cross sections for
baryon meson and baryon baryon systems we use the additive quark model(see 
Figure 11). The particles produced from such scatterings are determined by a 
multi-pomeron chain model using a Field-Feymann algorithm. For baryon($B$) 
anti-baryon($\overline{B}$) scattering and cross section, data is used for 
$N \overline{N}$ annihilation and elastic scattering. For annihilation
yields, we use a  flavor consistent meson meson multi-pomeron chain model.
For the rest of the yield a $B \overline{B}$ multi-pomeron chain model is
used. The elastic scattering obtained by this method is close to the data
for the $N \overline{N}$ system. 
   
\begin{figure}
\begin{center}
\mbox{
   \epsfysize 5.0in
   \epsfbox{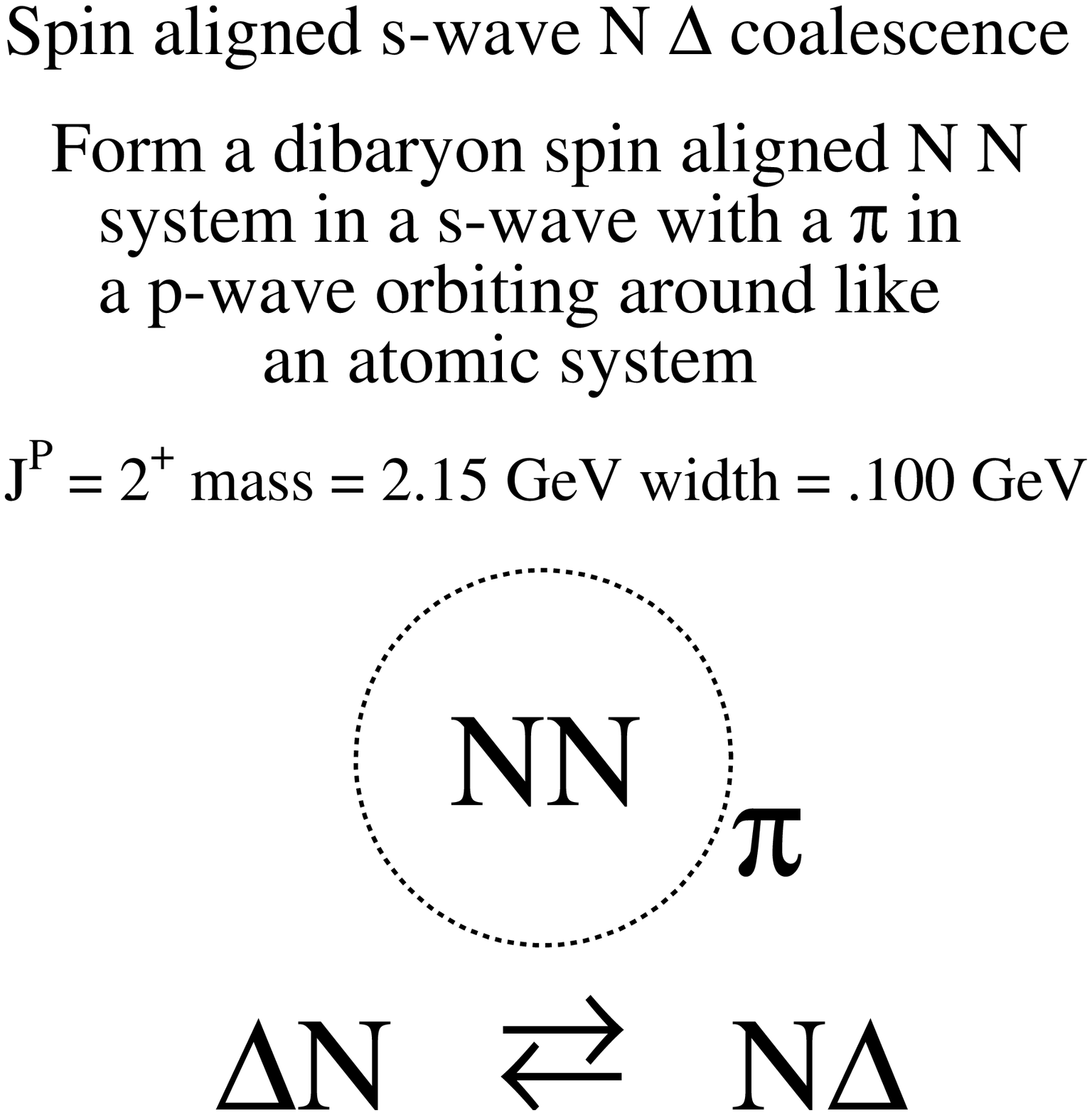}}
\end{center}
\vspace{2pt}
\caption{The dibaryon system mainly resonates in the s-wave $\Delta$ $N$ mode 
with a pion rotating in a p-wave about a spin aligned $N$ $N$ system which 
forms a isospin singlet. The pion moves back and forth forming $\Delta$ states 
with one nucleon and then the other.}
\label{fig3}
\end{figure}

\begin{figure}
\begin{center}
\mbox{
   \epsfysize 6.0in
   \epsfbox{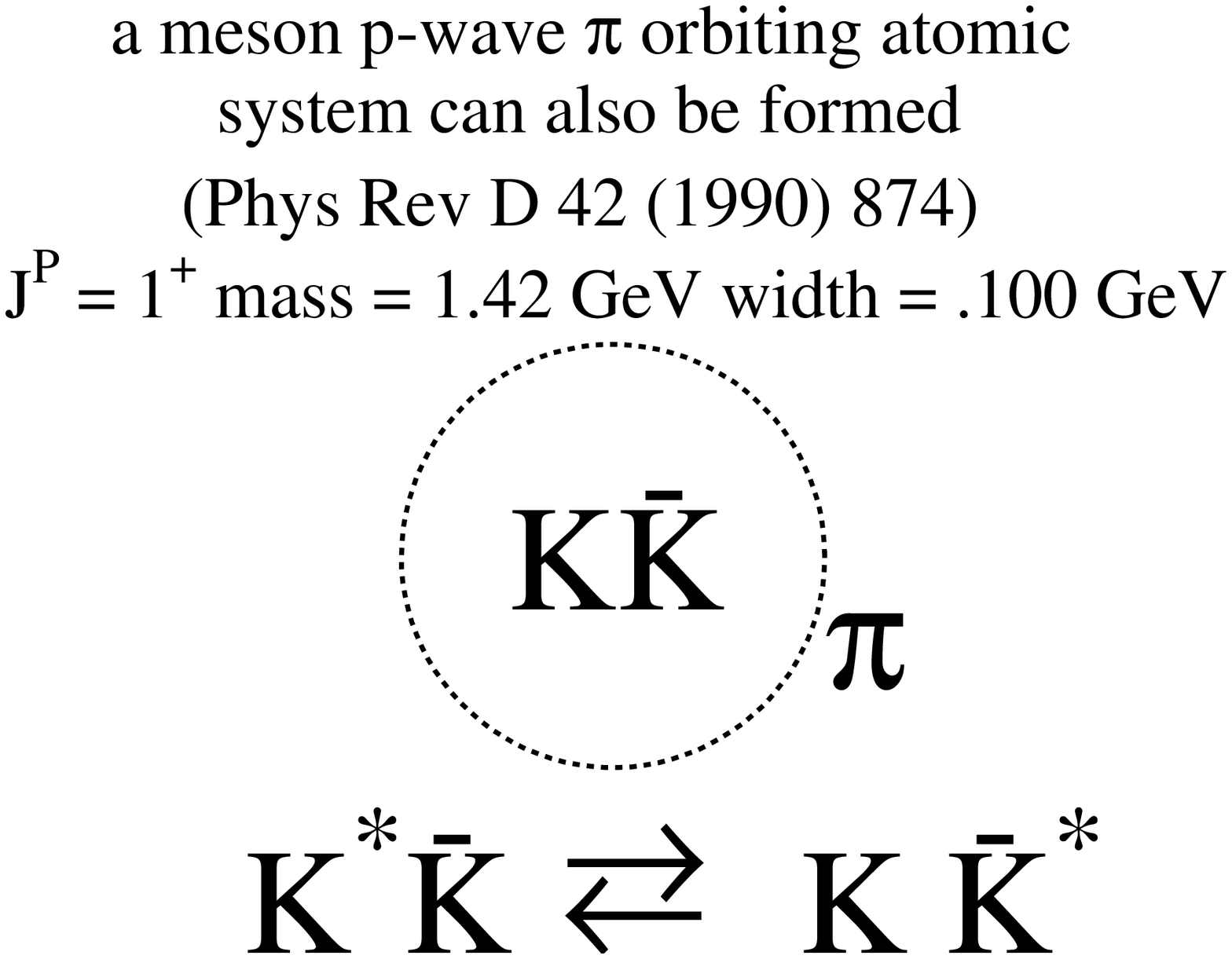}}
\end{center}
\vspace{2pt}
\caption{The meson system mainly resonates in the s-wave $K^*$ $\overline{K}$ 
and $K$ $\overline{K^*}$ mode with a pion rotating in a p-wave about a 
$K$ $\overline{K}$ system which forms a isospin triplet. The pion moves back 
and forth forming $K^*$ and $\overline{K^*}$ states with one $K$ or 
$\overline{K}$.}
\label{fig4}
\end{figure}

\begin{figure}
\begin{center}
\mbox{
   \epsfysize 6.0in
   \epsfbox{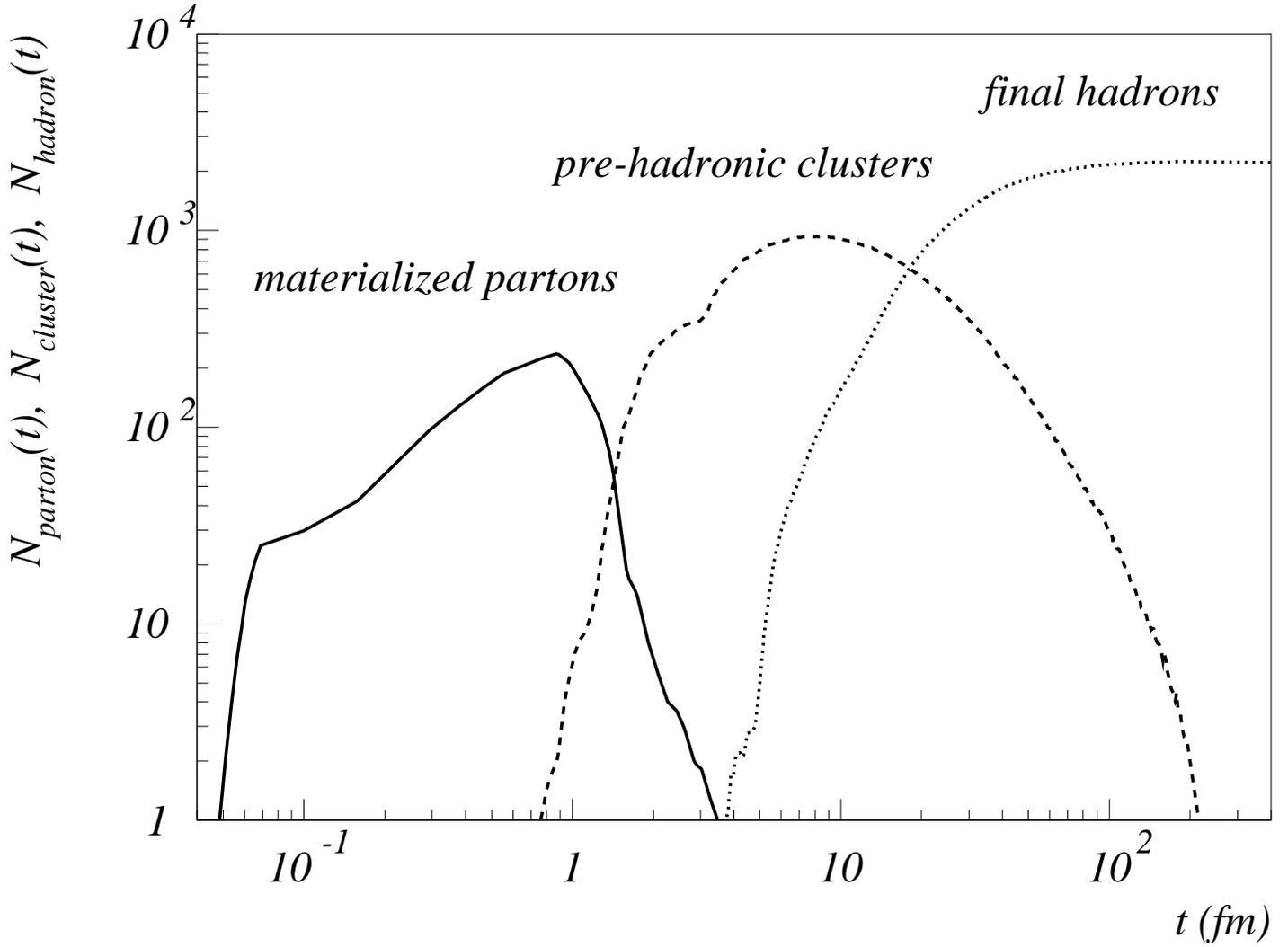}}
\end{center}
\vspace{2pt}
\caption{Time evolution of the total numbers of produced partons Np, 
pre-hadronic clusters Nc, and hadrons Nh during Au + Au collisions.
The time refers to the center-of-mass frame of the colliding nuclei.}
\label{fig5}
\end{figure}

\begin{figure}
\begin{center}
\mbox{
   \epsfysize 6.7in
   \epsfbox{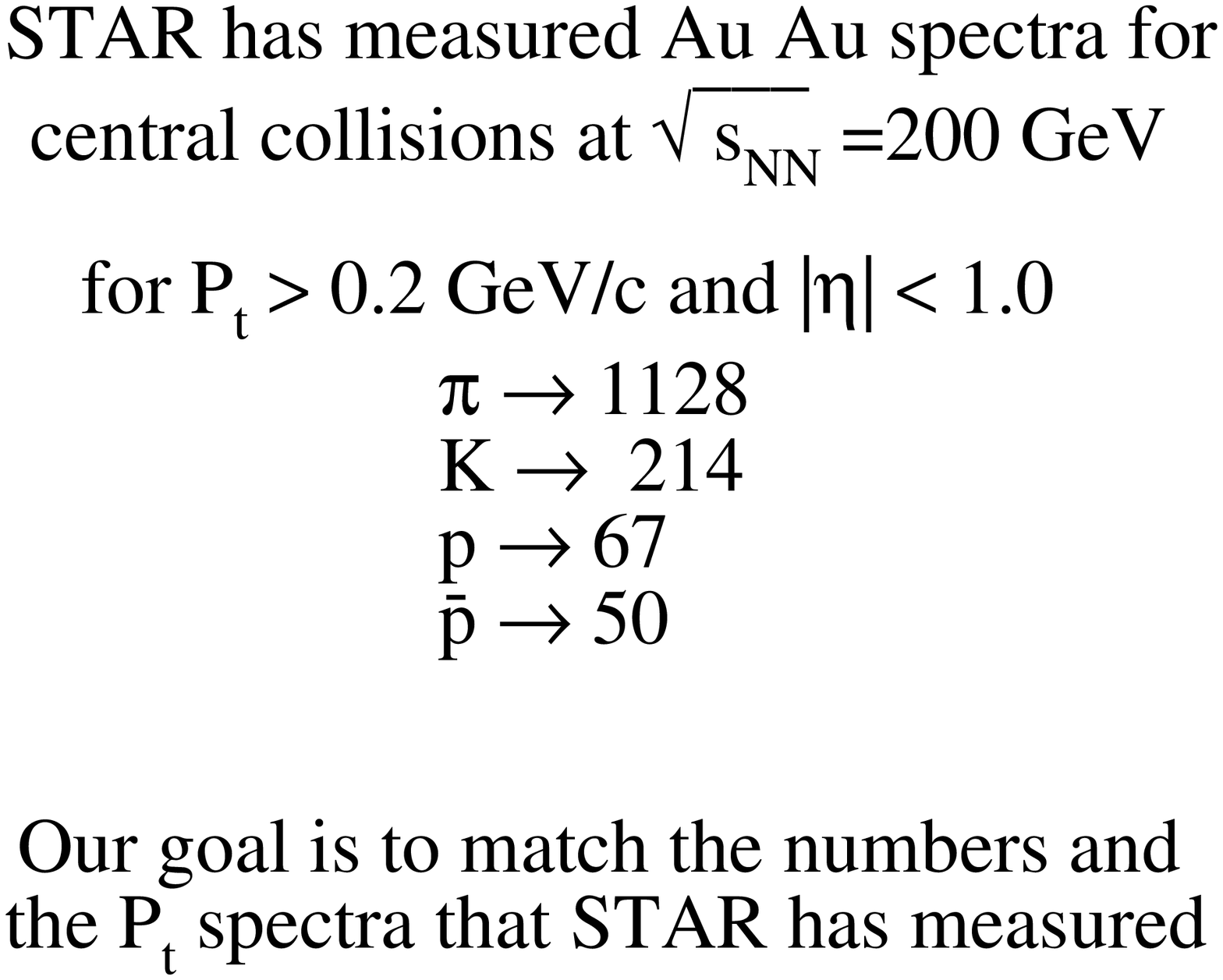}}
\end{center}
\vspace{2pt}
\caption{The measured Au+Au spectrum which we will tune for.}
\label{fig6}
\end{figure}

\begin{figure}
\begin{center}
\mbox{
   \epsfysize 7.9in
   \epsfbox{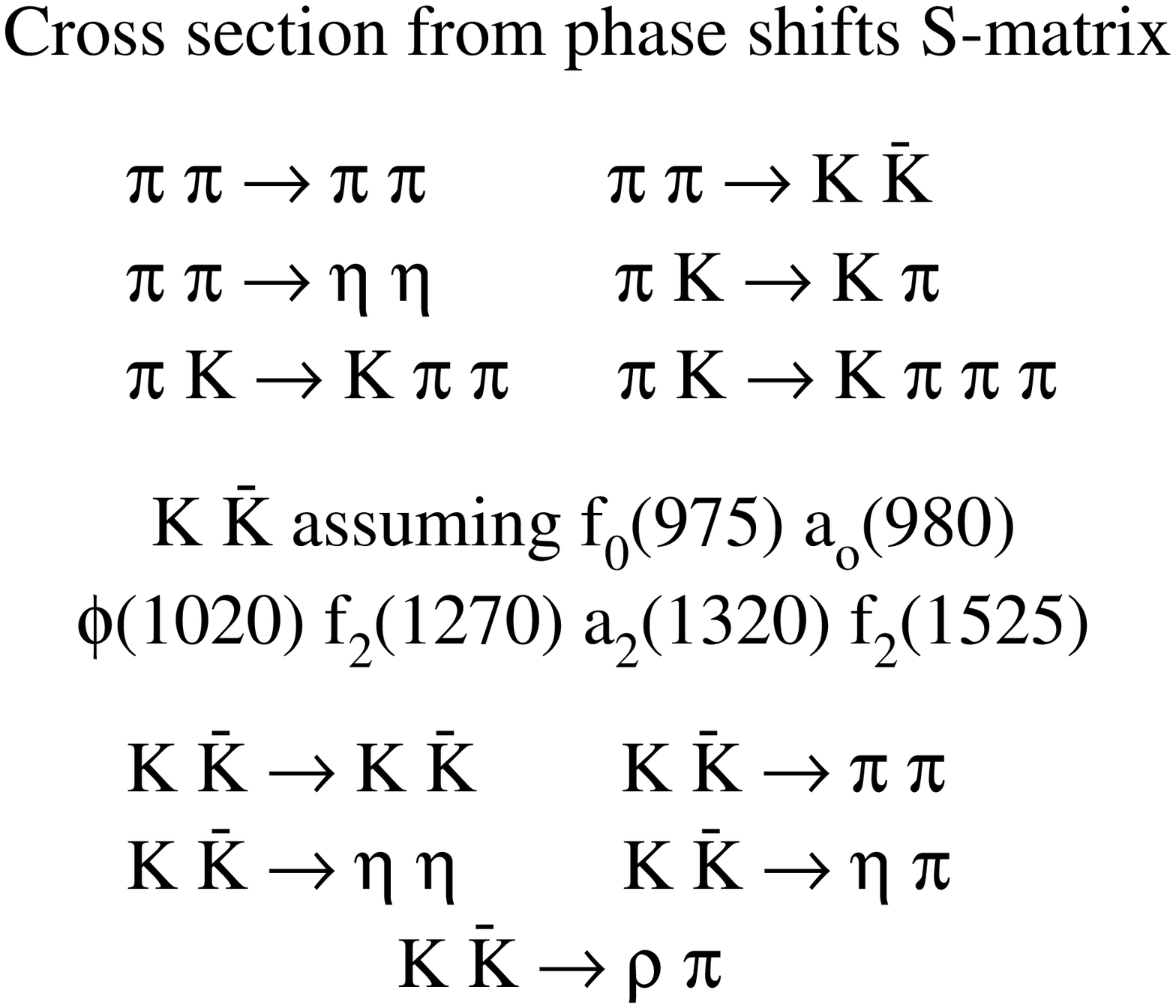}}
\end{center}
\vspace{2pt}
\caption{Cross sections for $\pi \pi$, $\pi K$, and $K \overline K$.}
\label{fig7}
\end{figure}

\begin{figure}
\begin{center}
\mbox{
   \epsfysize 7.9in
   \epsfbox{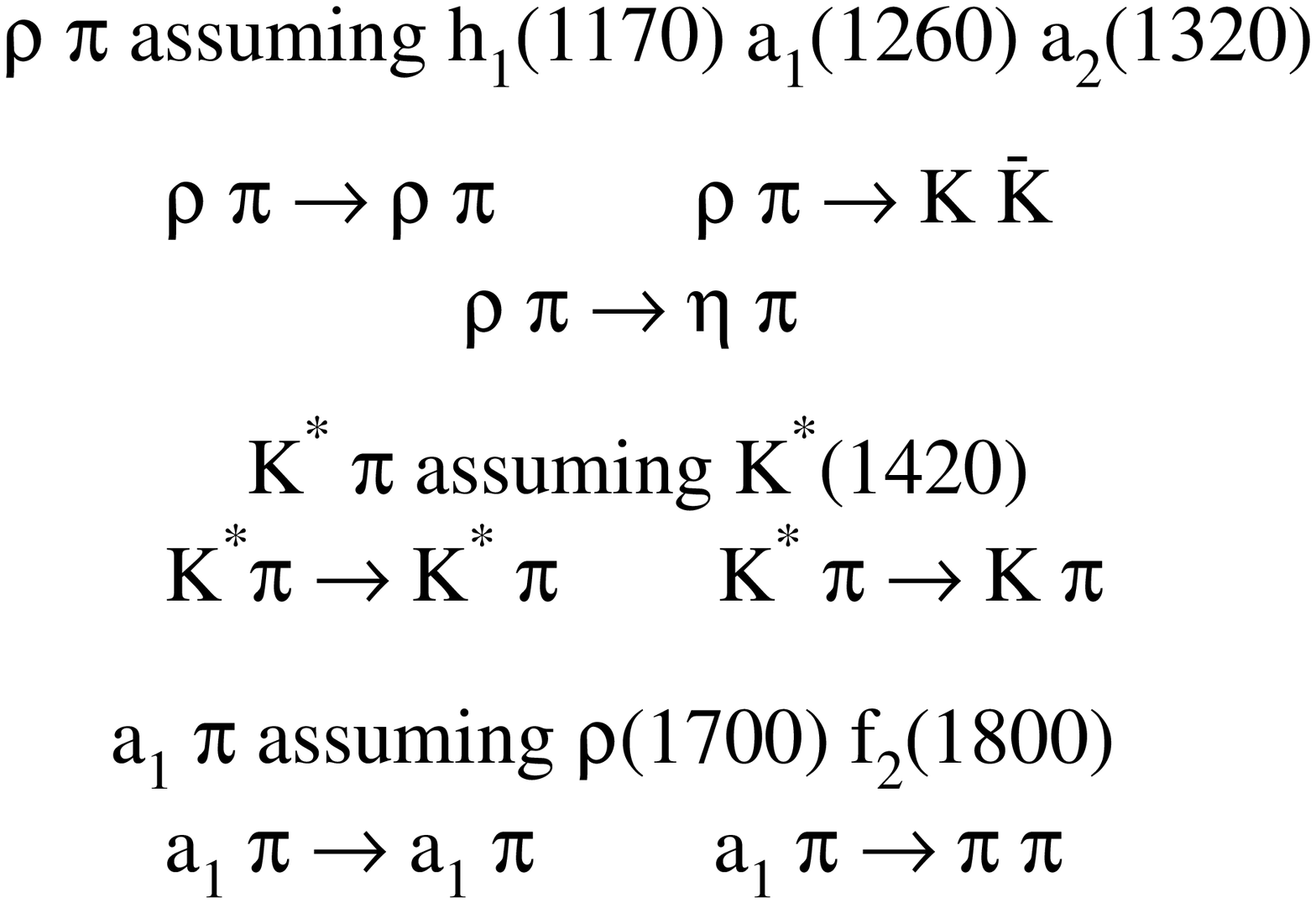}}
\end{center}
\vspace{2pt}
\caption{Cross sections for $\rho \pi$, $K^* \pi$, and $a_1 \pi$.}
\label{fig8}
\end{figure}

\begin{figure}
\begin{center}
\mbox{
   \epsfysize 7.9in
   \epsfbox{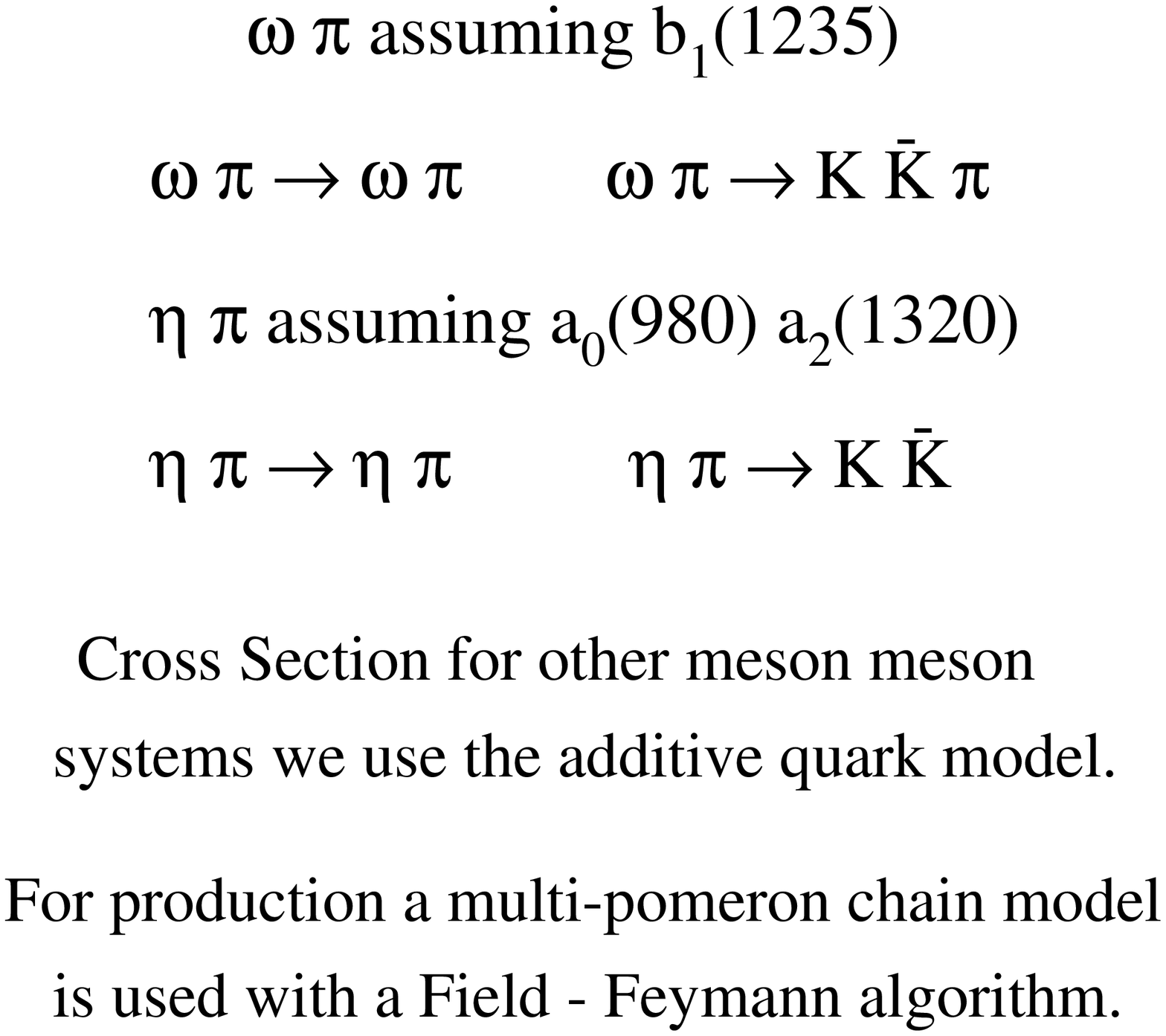}}
\end{center}
\vspace{2pt}
\caption{Cross sections for $\omega \pi$, $\eta \pi$, and others.}
\label{fig9}
\end{figure}

\begin{figure}
\begin{center}
\mbox{
   \epsfysize 7.9in
   \epsfbox{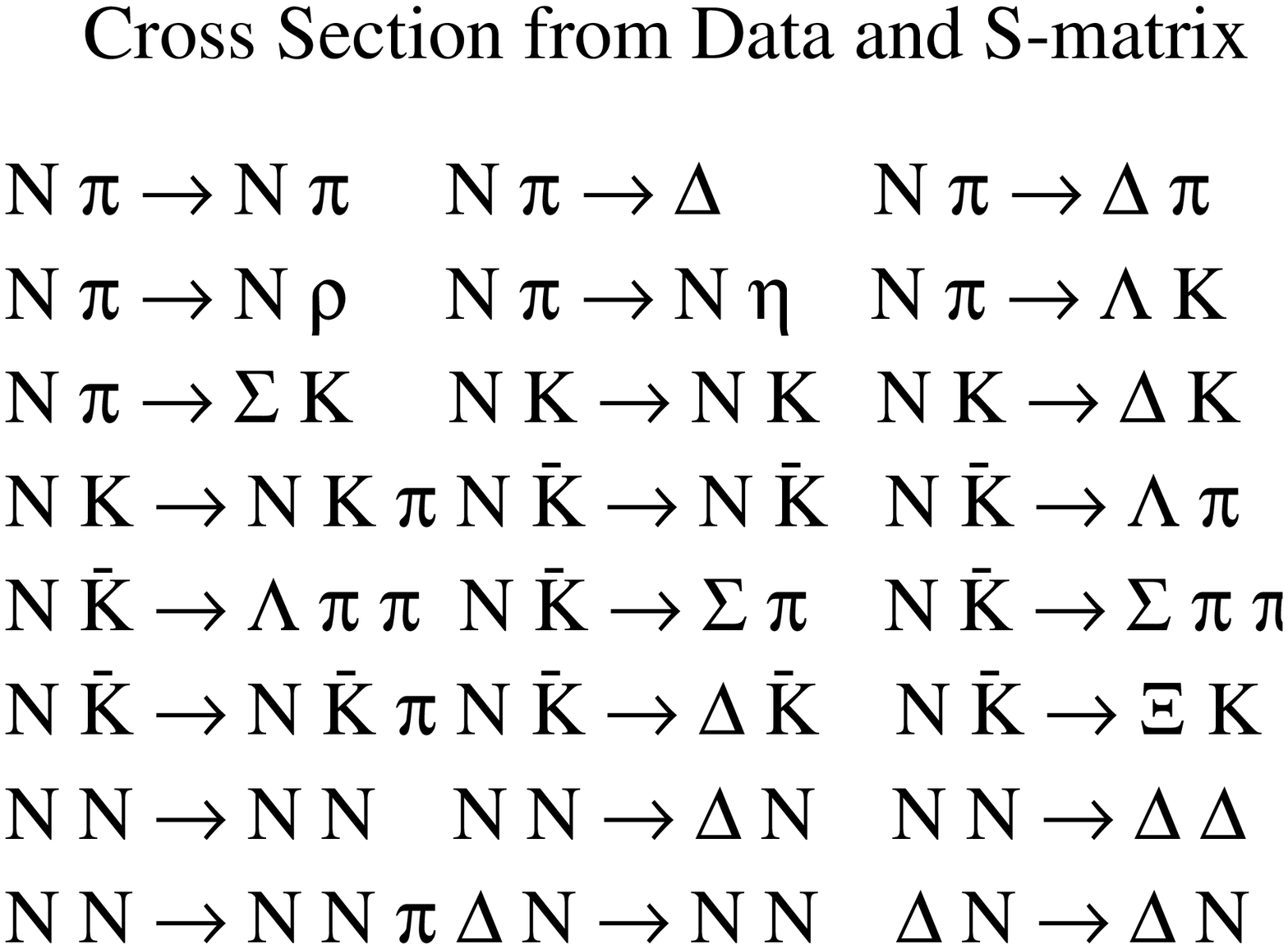}}
\end{center}
\vspace{2pt}
\caption{Cross sections for $N \pi$, $N K$, $N N$ and others.}
\label{fig10}
\end{figure}

\begin{figure}
\begin{center}
\mbox{
   \epsfysize 7.9in
   \epsfbox{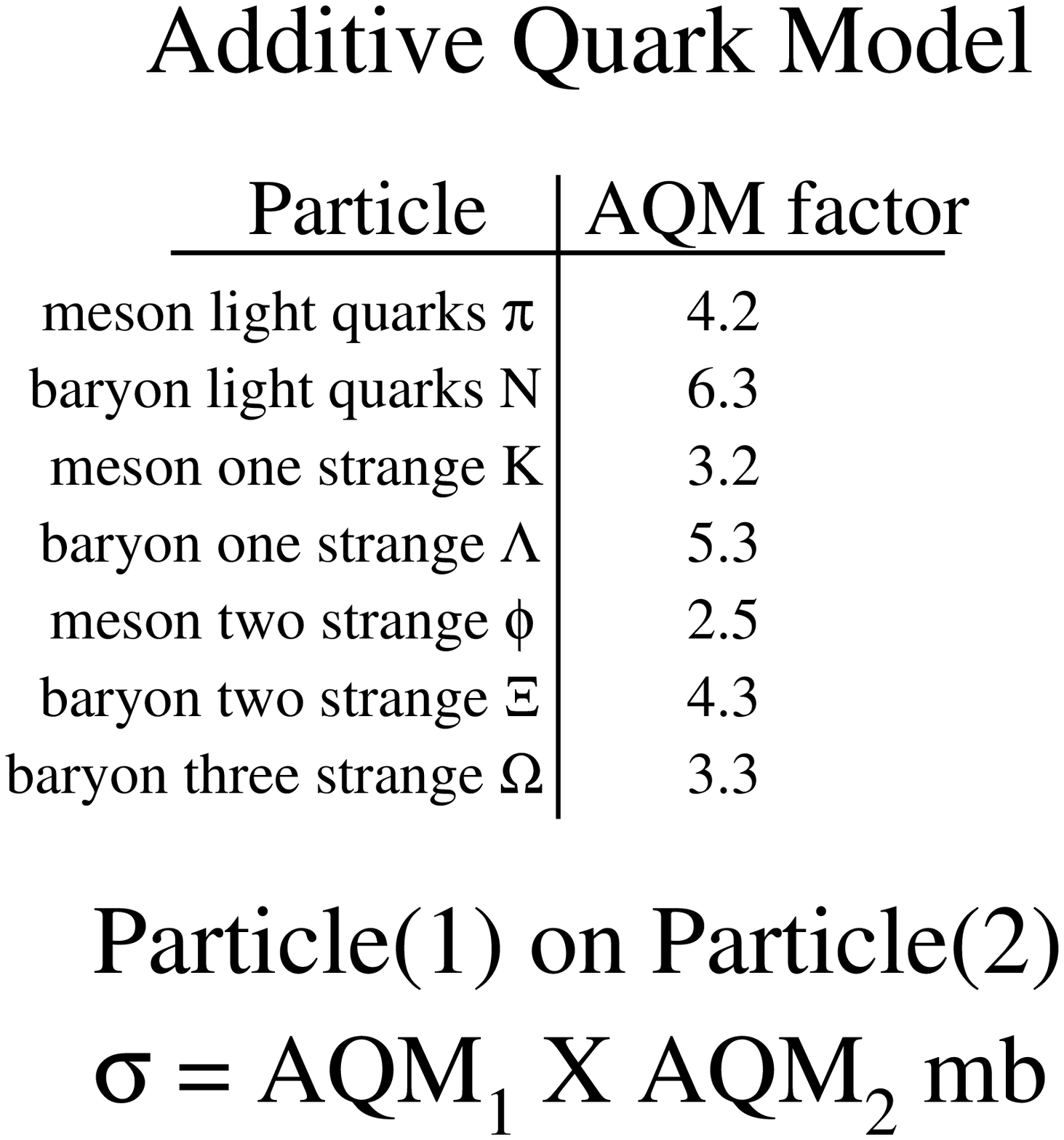}}
\end{center}
\vspace{2pt}
\caption{The additive quark model calculates the cross section for the 
scattering of any two particles based on a product of geometric factors.}
\label{fig11}
\end{figure}

\clearpage

The annihilation threshold effect is scaled to other $B \overline{B}$ 
scatterings using the $N \overline{N}$ ratios obtained in the above
algorithm.

We need to add the production of d's into the Monte Carlo code. Let us assume
that the formation of the $J^p = 2^+$ dibaryon state is the driving source of 
d's. We fit the d-wave $N N$ elastic scattering\cite{Arndt}, p-wave $d \pi$ 
elastic scattering\cite{Oh}, and p-wave $d \pi$ to d-wave $N N$\cite{Oh}. 
A three channel K-matrix was used to form a S-matrix, where the 
channels are d-wave $N N$, p-wave $d \pi$, and s-wave $\Delta N$ data. 
We are able to fit the above data if we use one K-matrix pole to generate the 
dibaryon 2.15 GeV state plus a far away pole and a flat none factorable 
background. Figure 12 shows the fit to elastic $N N$ scattering amplitude. 
Figure 13 shows the fit to $d \pi$ elastic scattering, while Figure 14 is the 
connection between $N N$ going to $d \pi$. 

The cross sections for $N$ $N$ $\rightarrow$ $\pi$ $d$, 
$\Delta$ $N$ $\rightarrow$ $\pi$ $d$, $\pi$ $d$ $\rightarrow$ 
$\Delta$ $N$ and $N$ $N$ where added to the hadron cascade part of the code. 
When we consider the known cross sections for $N N$ and $\Delta N$, the 
yield for charge pairs of $d \pi$ can be calculated and is plotted in
Figure 15. In our hadron cascade these scatterings are the only source of d's.
The production of d's and anti-d's is close to the values measured in 
Figure 1. The value of d's in the cascade would be much larger than the 
measured value except d's are destroyed by interacting with pions. Figure 16 
show the large $d \pi$ cross section of $\sim$ 250 mb. About 3/4 of these 
scattering remove the d's from the cascade.

 We achieve the yield and spectrum for Au+Au $\sqrt{s_{NN}}$=200 GeV central
collisions by adjusting the excited hadrons in our cylinder of radius 10.0 
Fermi. We generate enough events at $\sqrt{s_{NN}}$=200 GeV central Au+Au 
collisions in order to obtain 1 million $d$ or $\overline{d}$ events in the 
STAR acceptance. Out of the 1 million events there were 230,000 pairs of either
$d$ $\pi$ or $\overline{d}$ $\pi$ which decayed in the STAR acceptance. 
The effective mass distribution of these pairs are plotted as solid points 
in Figure 17.

In order to obtain the mass spectrum from the data, We need to determined the 
uncorrelated background of either $d$ or $\overline{d}$ paired with a charge 
particle in a average event. For each of the 1 million events we can pair up 
either the $d$ or $\overline{d}$ with all charge particles(which then is 
assumed to be a pion) in that event and plot the total number of pairs as a 
function of effective mass. From this pair spectrum we then subtract the 
average uncorrelated spectrum times the number of events. We can determine this
average uncorrelated spectrum by mixed event methods taking the same $d$ and 
$\overline{d}$ paired with the charged particles from other events. The
subtracted effective mass spectrum is the open points of Figure 17.
We see that we have recovered the mass spectrum.

\section{Summary and Discussion}

In the first section of this manuscript we consider baryons and anti-baryons 
up to a baryon number five. These states decayed by the weak interaction. The 
exotic states that decay strongly is considered in the second section. In 
order to develop methods for such research we consider a dibaryon(2.15) 
$J^P$ = $2^+$ state which decays into d $\pi$. 

\begin{figure}
\begin{center}
\mbox{
   \epsfysize 8.5in
   \epsfbox{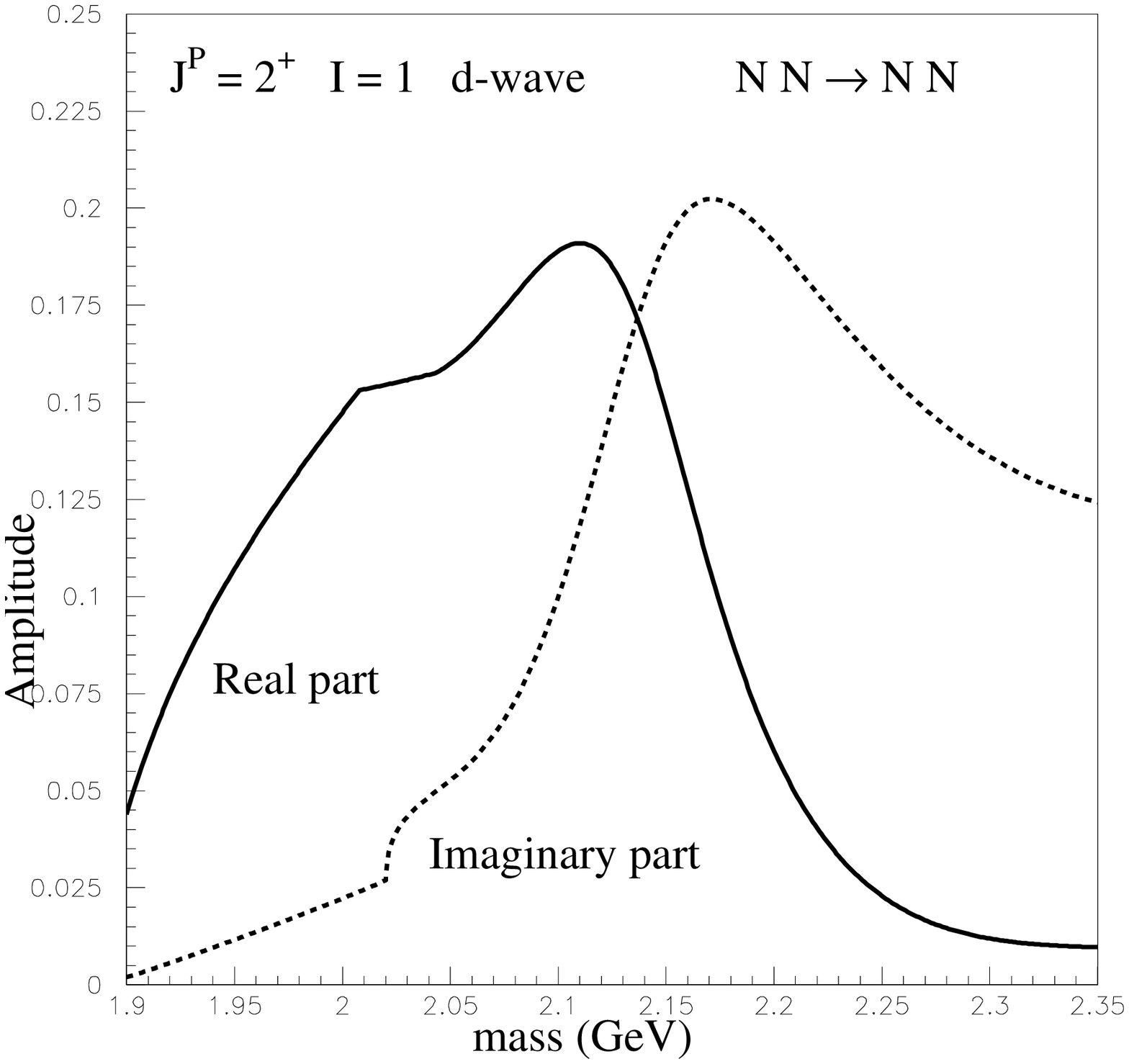}}
\end{center}
\vspace{2pt}
\caption{The real and imaginary parts of the elastic scattering T-matrix
amplitude for $N N \rightarrow  N N$ as a function of mass in GeV.}
\label{fig12}
\end{figure}

\begin{figure}
\begin{center}
\mbox{
   \epsfysize 8.5in
   \epsfbox{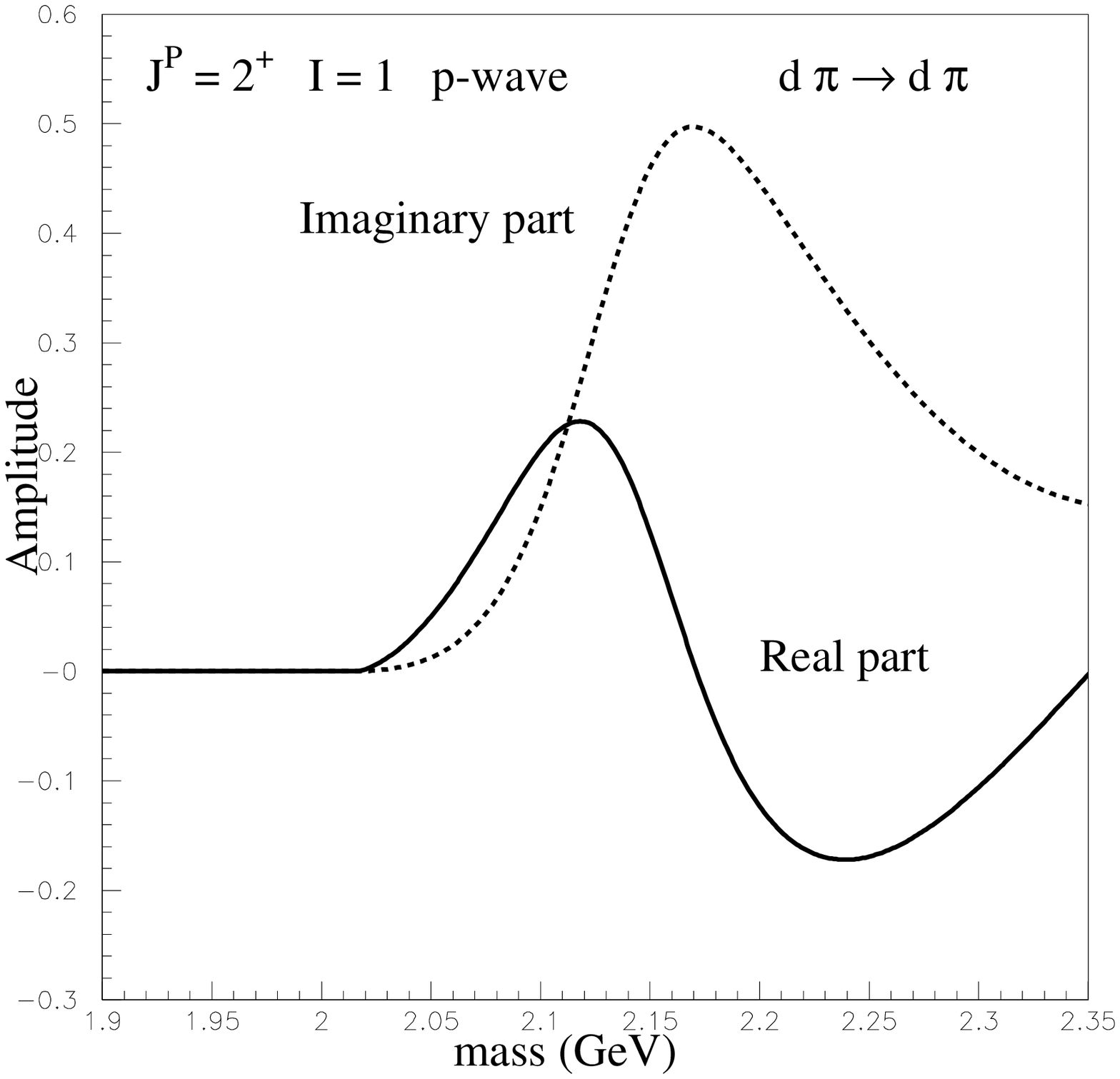}}
\end{center}
\vspace{2pt}
\caption{The real and imaginary parts of the elastic scattering T-matrix
amplitude for $d \pi \rightarrow  d \pi$ as a function of mass in GeV.}
\label{fig13}
\end{figure}

\begin{figure}
\begin{center}
\mbox{
   \epsfysize 8.5in
   \epsfbox{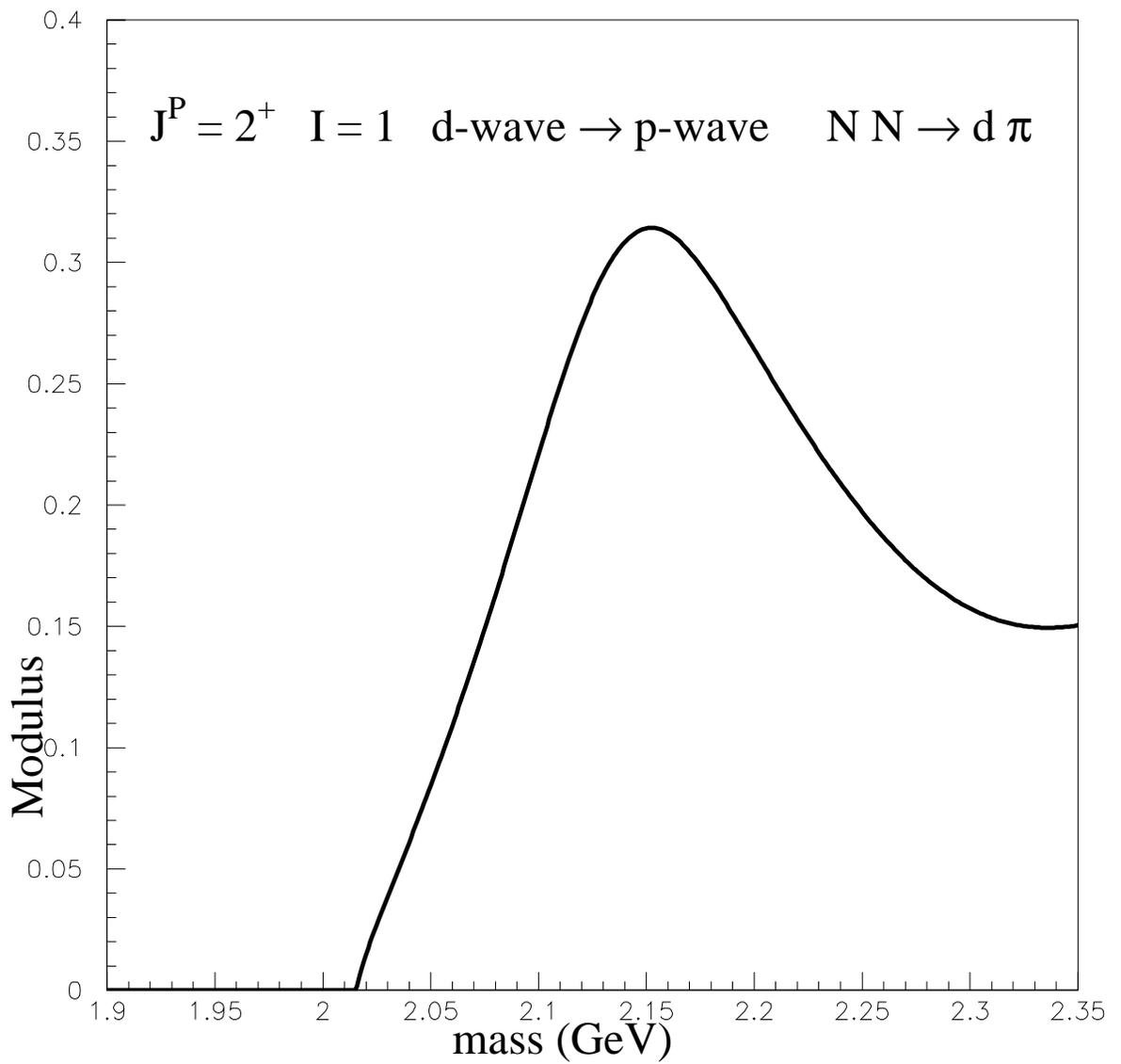}}
\end{center}
\vspace{2pt}
\caption{The modulus of the inelastic scattering T-matrix amplitude for $N N
\rightarrow d \pi$ as a function of mass in GeV.}
\label{fig14}
\end{figure}

\begin{figure}
\begin{center}
\mbox{
   \epsfysize 8.5in
   \epsfbox{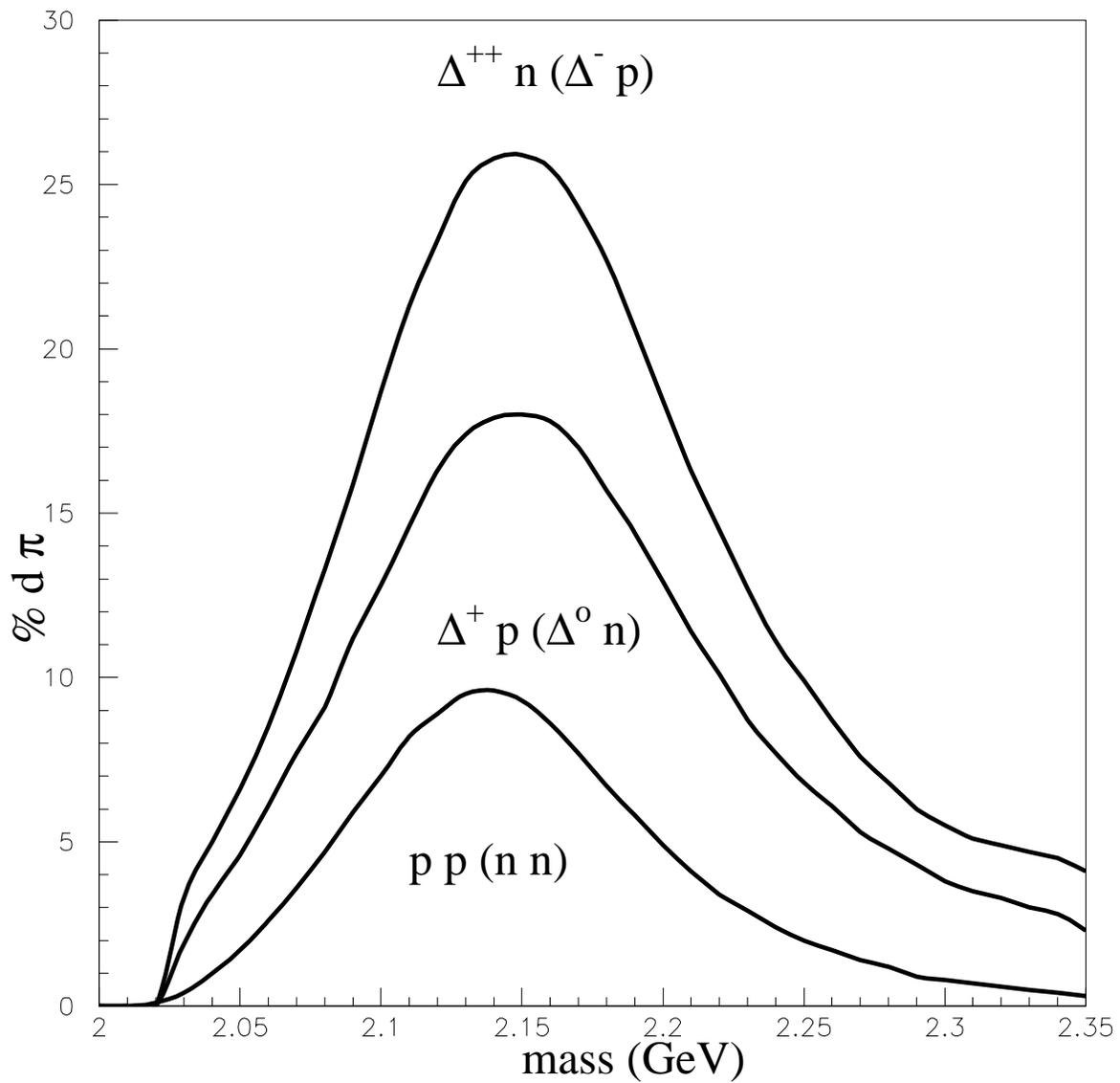}}
\end{center}
\vspace{2pt}
\caption{The percentage of $d \pi$ charge pairs produced in $N N$ and 
$\Delta N$ scattering as a function of mass in GeV.}
\label{fig15}
\end{figure}

\begin{figure}
\begin{center}
\mbox{
   \epsfysize 8.5in
   \epsfbox{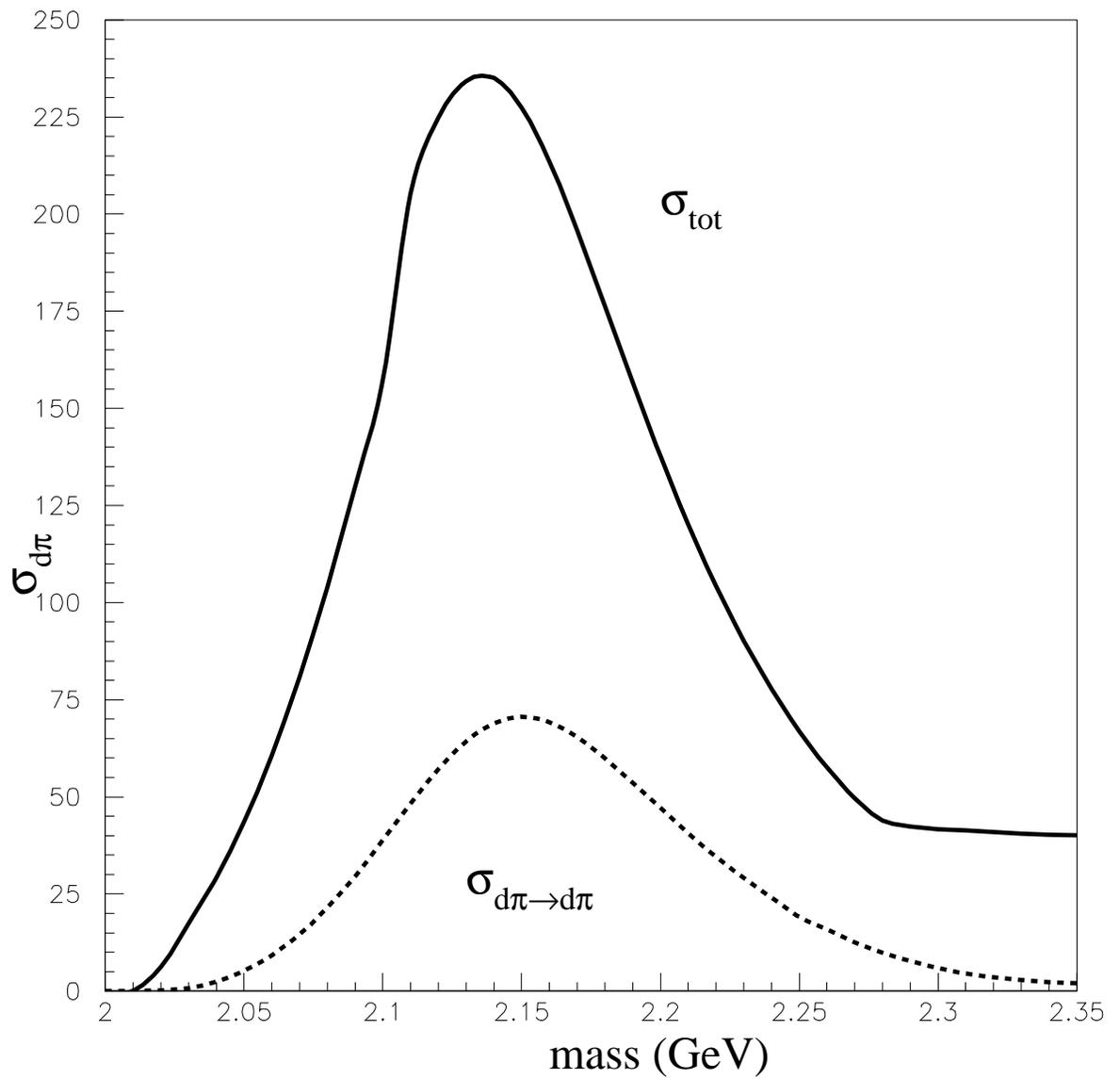}}
\end{center}
\vspace{2pt}
\caption{The $d \pi$ total and elastic cross section in millibarns(mb) as a 
function of mass in GeV.}
\label{fig16}
\end{figure}

\begin{figure}
\begin{center}
\mbox{
   \epsfysize 6.0in
   \epsfbox{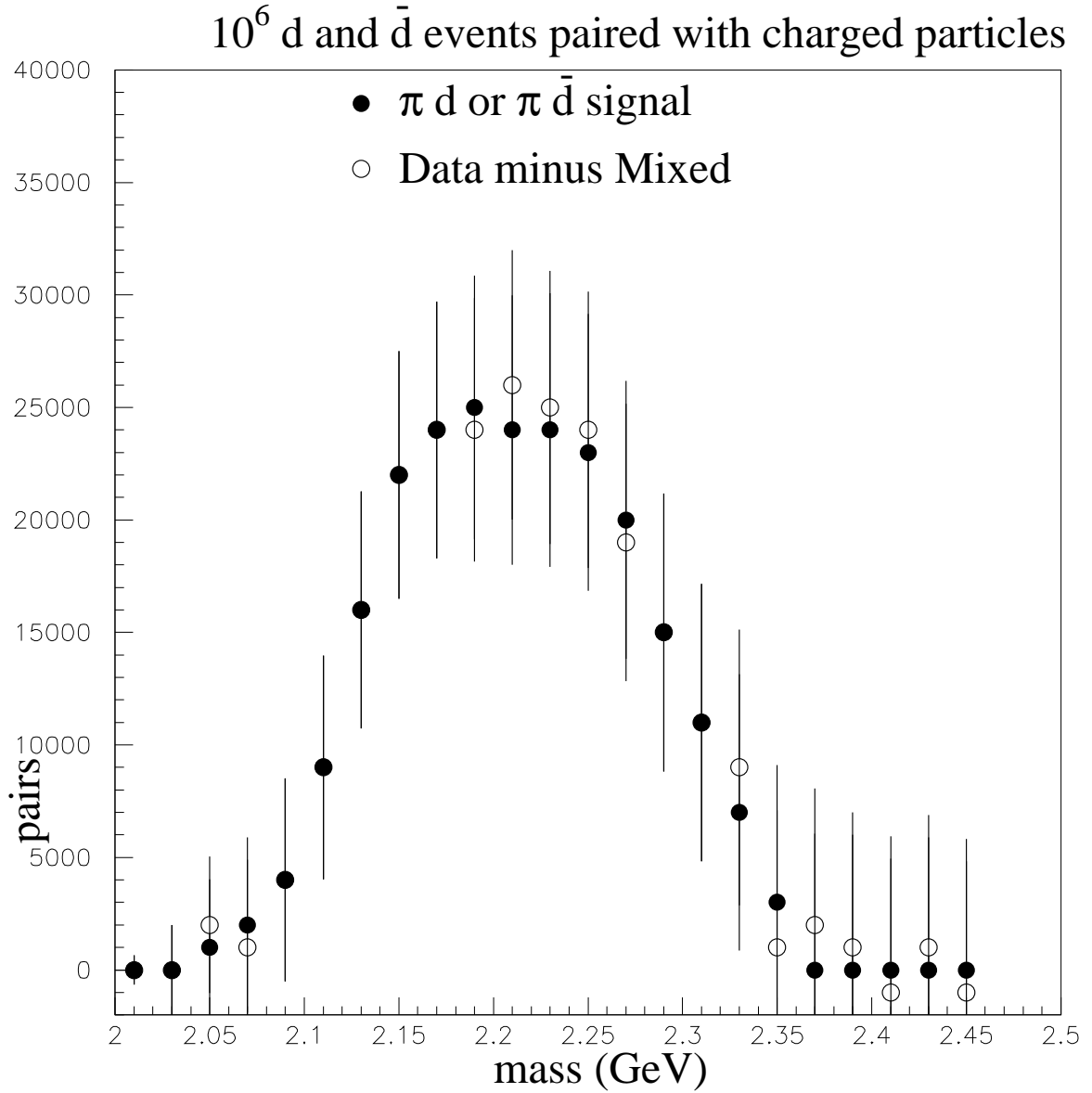}}
\end{center}
\vspace{2pt}
\caption{The number of $d$ or $\overline{d}$ paired with charged pions coming
from $10^6$ dibaryons decays within the STAR acceptance plotted as solid 
points. The open points are form by all $d$ and $\overline{d}$ paired with
the charged particles in each event in the star acceptance minus the same
$d$ and $\overline{d}$ paired with the charged particles from other 
events(mixed events).}
\label{fig17}
\end{figure}

\clearpage

We create a Monte Carlo simulation that should give realistic events structure 
with realistic dibaryon production. With the ability to measure hundreds of 
million ultra-relativistic heavy ion collisions, we predicted that a clear 
dibaryon signal decaying into $d \pi$ should be measured.

\section{Acknowledgments}

This research was supported by the U.S. Department of Energy under Contract No.
DE-AC02-98CH10886.


\begin{thebibliography}{99}
\bibitem{Armstrong} T.A.~Armstrong {\it et al.}, Phys. Rev. Lett. 83 (1999) 
5431.
\bibitem{He4} H.~Agakishiev {\it et al.}, Nature 473 (2011) 353.
\bibitem{Arndt} R.A.~Arndt {\it et al.}, Phys. Rev. C 76 (2007) 025209.
\bibitem{Oh} C.H.~Oh {\it et al.}, Phys. Rev. C 56 (1997) 635.
\bibitem{Schiff} D.~Schiff and J.~Tran Thanh Van, Nucl. Phys. B5 (1968) 529.
\bibitem{Longacre} R.~Longacre, Phys. Rev. D 42 (1990) 874.
\bibitem{Klaus} K.~Geiger and R.~Longacre, Heavy Ion Phys. 8 (1998) 41.
\end{thebibliography}
\end{document}